\documentclass[revtex4]{emulateapj}
\tightenlines

\begin{document}

\title{Compact dusty clouds in cosmic environment}
\author{V. N. Tsytovich$^{1,2}$, A. V. Ivlev$^2$, A. Burkert$^2$, G. E.~Morfill$^2$}
\email[E-mail:~]{ivlev@mpe.mpg.de}
\affiliation{$^1$General Physics Institute, Russian Academy of Sciences, 117942 Moscow, Russia\\
$^2$Max-Planck-Institut f\"{u}r extraterrestrische Physik, 85741 Garching, Germany}

\begin{abstract}
A novel mechanism of the formation of compact dusty clouds in astrophysical environments is discussed. It is shown that the
balance of collective forces operating in space dusty plasmas can result in the effect of dust self-confinement, generating
equilibrium spherical clusters. The distribution of dust and plasma density inside such objects and their stability are
investigated. Spherical dusty clouds can be formed in a broad range of plasma parameters, suggesting that this process of
dust self-organization might be a generic phenomenon occurring in different astrophysical media. We argue that compact dusty
clouds can represent condensation seeds for a population of small-scale, cold, gaseous clumps in the diffuse interstellar
medium. They could play an important role in regulating its small-scale structure and its thermodynamical evolution.
\end{abstract}

\keywords{plasmas -- ISM: clouds -- ISM: dust -- ISM: structure}

\maketitle

\section{Introduction}
\label{intro}

Self-organization of dusty (complex) plasmas has been observed in numerous experiments. Different types of structures formed
in dusty plasmas under microgravity conditions (in experiments performed on the International Space Station) as well as on
ground include compact clusters, voids surrounded by dust shells, vortices, etc. \citep{UFN,Fortov05,TsytoBook,Bonitz10}.
The mechanisms governing such phenomena are associated with plasma fluxes generated due to electron and ion absorption on
grains. Plasma fluxes exert the forces which can result in the effect of dust self-confinement.

It was shown theoretically that homogeneous dusty plasmas are intrinsically unstable \citep{MT00,BT01,TW03}. For small dusty
clouds the long-range ($\propto r^{-1}$) attraction between grains can be caused by ion ``shadowing'' forces, induced due to
plasma absorption on the grain surfaces \citep{TsytoBook}. The theory of such gravitation-like instability was developed for
laboratory conditions, and the analogy with the Jeans instability was pointed out. The important difference between the
laboratory and space conditions are \citep{UFN,WhittetBook,Draine09}: (i) the volume ionization in astrophysical environment
is much less important then in laboratory plasmas, (ii) the grain screening in space is described (to a very good accuracy)
by a simple linear Debye screening, while in laboratory conditions it is normally highly nonlinear, and (iii) the ratio of
ion to electron temperatures in space is close to unity, while in laboratory experiments it is typically $\sim10^{-2}$.
These distinctions require new theoretical and numerical treatment of self-organization in space dusty plasmas.

The principal aim of this paper is to point out the importance and possibility of dust self-organization in cosmic
environments like the diffuse interstellar medium (ISM). We present and solve the basic set of self-consistent equations
describing equilibrium dusty clouds, identify necessary conditions for such clouds to exist, and analyze their stability.
Unlike stars, where the equilibrium is primarily governed by gravity and pressure, the dusty clouds are formed due to the
balance of the ion drag force (associated with the self-consistent plasma flux) and the electrostatic force on charged
grains.

We show that cosmic dust can form stable spherical clouds, with typical sizes of the order of 10-100~AU or less and with
total mass of the order of $10^{-3}$ Earth mass or below. Although the dust density inside the clouds can exceed the ambient
density by many orders of magnitude -- therefore we call them ``compact dusty clouds'' -- they remain optically thin. We
predict that this process can occur in a broad range of plasma parameters, which indicates that such self-organization might
be a generic phenomenon operating in different astrophysical media.

One particularly interesting application is the formation of compact dusty clouds in the diffuse ISM. Ultra-high resolution
observation of interstellar absorption \citep{BK05,S13,C13} provide evidence that the diffuse ISM is structured on scales
below 1 pc, indicating the presence of tiny, distinct, dense and cold HI cloudlets, the origin of which is not understood up
to now. This gas component might play an important role in regulating the thermodynamical state of the ISM and in triggering
phase transitions.

The paper is organized as follows: In Sec.~\ref{assumptions} we summarize principal simplifying assumptions relevant to
interstellar environment; in Sec.~\ref{processes} we discuss the generic mechanism resulting in formation of compact dusty
clouds, and also introduce proper normalization of variables; in Sec.~\ref{master_eqs} we assume negatively charged dust and
formulate self-consistent equations describing equilibrium spherical clouds in this case; in Sec.~\ref{equilibrium_results}
we solve the equations numerically, to present distributions of parameters inside the clouds and estimate their major
characteristics (such as size, mass, and dust density) for some idealized ISM phases; in Sec.~\ref{stability} we analyze
stability of the obtained equilibrium clouds; in Sec.~\ref{positive} we consider a simplified model for positively charged
dust and show that equilibrium compact clouds can be formed in this case as well; in Sec.~\ref{diffuse} we discuss the
effect of compact dusty clouds on the small-scale structure of the diffuse ISM; and in Sec.~\ref{conclusions} we summarize
the results and discuss possibilities to observe compact dusty clouds.

\section{Idealized astrophysical environments and simplifying assumptions}
\label{assumptions}

Table~\ref{tab1} summarizes physical parameters for some idealized ISM phases \citep{WhittetBook,Draine98,Yan04}: the
reflection nebula (RN), cold neutral medium (CNM), and warm neutral medium (WNM). The principal parameters relevant for the
further analysis are the gas temperature $T$, the atomic hydrogen density $n_{\rm H}$, the molecular hydrogen density
$n_{{\rm H}_2}$, and the density of hydrogen ions $n_{{\rm H}^+}$. (Below, unless explicitly specified, we shall employ the
notation $n_i$ for the ion density and $n_n$ for the total density of neutrals.) For these conditions we can make the
following simplifying assumptions:

{\it I. Dust grains have the same size.} Dust in astrophysical environments is extremely polydisperse. In the range between
several nm to a few tenths of $\mu$m, the dust size distribution can normally be approximated by the model MRN dependence
\citep{Mathis}, $dn_d/da\propto a^{-3.5}$. As we will show in the next section, the two major forces -- ion drag and
electrostatic -- whose balance results in the formation of equilibrium dusty clouds are dominated by the small-size part of
the distribution. Therefore, we can restrict our analysis to some effective size (radius of the effective small-size
cutoff). In the normalized (dimensionless) form, the governing equations used for the analysis {\it do not contain the
size}, so that general results obtained below remain unaffected by this assumption.

\begin{table}
\caption{Idealized ISM Phases} \label{tab1} \centering
\begin{tabular}{lcccc} \hline\hline\\
{Parameter} & {RN} & CNM & WNM  \\[1mm]
\hline\\
$T$~(K)                       & 100        & 100       & 6000        \\[1mm]
$n_{\rm H}~({\rm cm}^{-3})$   & $10^{3}$   & 30        & 0.4         \\[1mm]
$2n_{{\rm H}_2}/n_{\rm H}$    & $10^{-2}$  & 0         & 0           \\[1mm]
$n_{{\rm H}^+}/n_{\rm H}$     & $10^{-3}$  & $10^{-3}$ & $10^{-1}$   \\[1mm]
\hline\hline\\
\end{tabular}
\end{table}


{\it II. Linear regime of dust screening.} The criterion of linearity of the plasma-dust interaction is very well satisfied
for astrophysical conditions, which allows us to assume a linear Debye screening of dust \citep{TsytoBook}. For a grain of
charge $eZ$ and radius $a$, this criterion requires the potential energy of the ion-grain interaction at the Debye length
$\lambda_{\rm D}$ to be much smaller than the plasma temperature, i.e., $|Z|e^{2}/\lambda_{\rm D}k_{\rm B}T\sim
a/\lambda_{\rm D}\ll 1$. This assumption significantly simplifies the force balance equation.

{\it III. Fluid description of ions.} We assume that the ion collisions with neutrals as well as with charged grains are
frequent enough, so that the resulting mean free path of ions is much smaller than any macroscopic length scale of the
problem. This allows us to introduce the local friction force on ions, which is proportional to their local flow velocity.
Furthermore, we assume the flow velocity to be much smaller than the ion thermal velocity (which is confirmed a posteriori
by the obtained solutions).

{\it IV. External forces and background inhomogeneities are negligible.} The analysis below shows that typical sizes of
dusty clouds are relatively small ($\lesssim30$~AU). Therefore, neglecting the centrifugal forces and inhomogeneities due to
the MHD turbulence \citep{Yan04,Yan03} is a reasonable approximation. The condition to neglect tidal forces is discussed in
Appendix~\ref{app1}.

These assumptions are only aimed to make the physical picture as transparent as possible, and demonstrate that the plasma
absorption on dust grains results in a universal self-confinement effect which can lead to the formation of stable compact
dusty clouds in cosmic environment.

\section{Physical processes important for cloud formation}
\label{processes}

\begin{figure}
\includegraphics[width=\columnwidth,clip=]{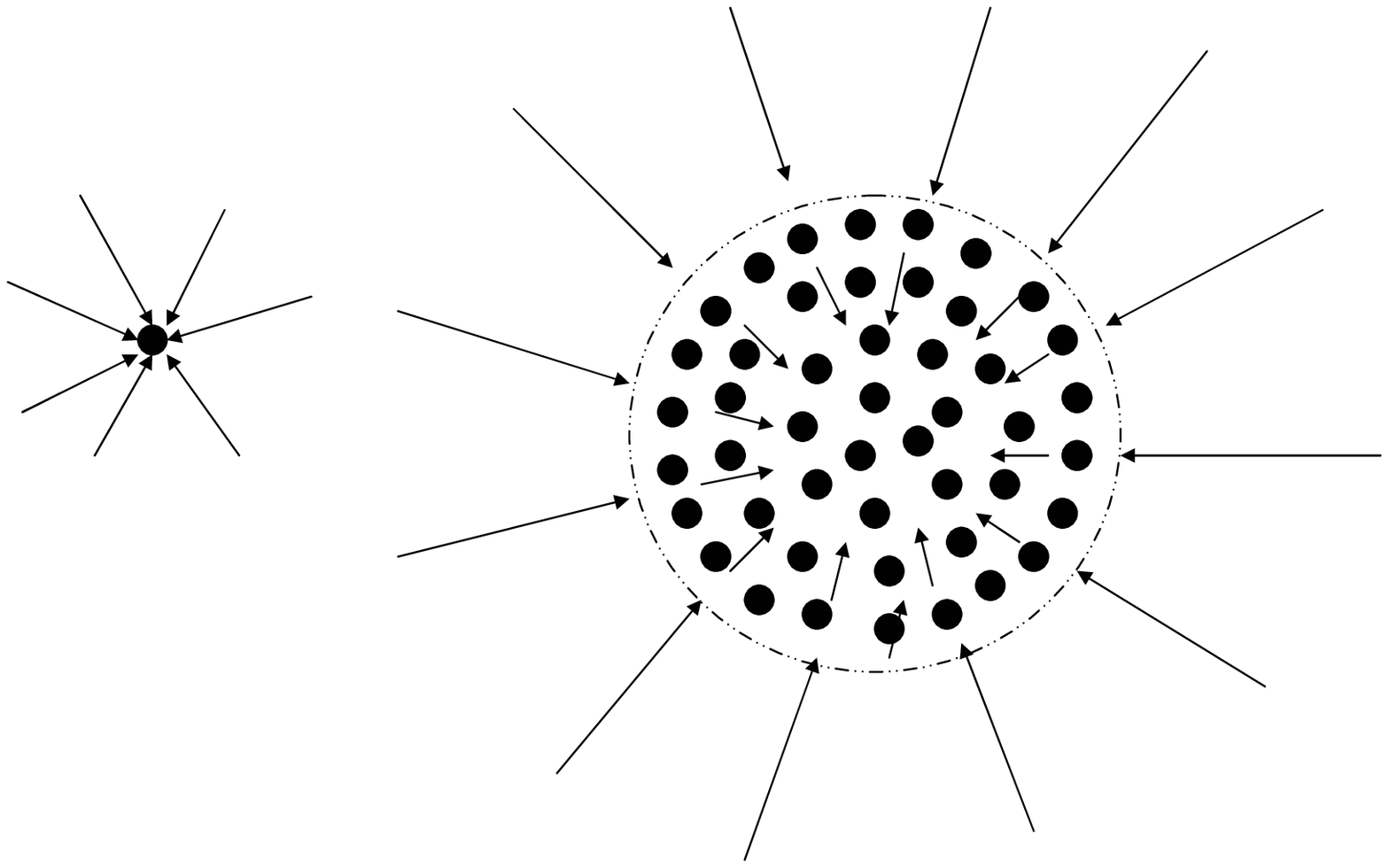}
\caption{Sketch showing (left) a {\it charging} flux on a single dust grain and (right) a {\it global} flux induced by a
spherical dusty cloud.} \label{Fig.1}
\end{figure}

Figure~\ref{Fig.1} illustrates the principal role of plasma fluxes for dusty clouds.

The absorption of electrons and ions by a single grain leads to its charging. In a plasma where external UV radiation does
not play the dominant role the resulting charge is {\it negative}, because electrons are much faster than ions. Then the
equilibrium charge $-eZ$ of a grain of radius $a$, being presented in the dimensionless form \citep{Fortov05},
\begin{equation}\label{norm_charge}
z=\frac{e^{2}Z}{ak_{\rm B}T},
\end{equation}
is normally about a few (for isothermal hydrogen plasmas $z\simeq2.5$). In this paper we mostly discuss this charging
regime. The case when UV radiation dominates and grains are charged {\it positively} is considered in Sec.~\ref{positive}.

For a large dusty cloud, in addition to the local electron and ion fluxes in a close proximity of each grain, there is also
a global plasma flux pointed inwards. The total flux ${\bf J}$ fluctuates around its equilibrium value, ${\bf J}=\langle{\bf
J}\rangle +\delta{\bf J}$, both due to fluctuations of plasma fluxes on individual particles and fluctuations of the
particle density (high-frequency fluctuations in electron-ion plasmas, not associated with dust and hence irrelevant to our
problem, are neglected). The plasma flux is inevitably accompanied by the self-consistent electric field, ${\bf
E}=\langle{\bf E}\rangle +\delta {\bf E}$. The fluctuating field (associated with the fluctuating flux and/or grain charges)
can result in a variety of interesting effects, such as stochastic particle acceleration -- also in the astrophysical
environment \citep{IvlevApJ,Hoang12}.

In this paper we neglect the fluctuations and focus on the effect of the regular components (so that the notation
$\langle\ldots\rangle$ is omitted below). According to Fig.~\ref{Fig.1}, the radial flux (due to absorption in a spherical
dusty cloud) is always pointed towards the cloud center. The radial electric field generated in this case is directed
inwards when grains are charged negatively (while for positively charged grains it is pointed outwards, see
Sec.~\ref{positive}). Thus, the average flux provides the confining effect via the ion drag force, and the average electric
field tends to compensate for it via the electrostatic force. Let us elaborate on this mechanism.

The electric field force acting on a single charged grain is
\begin{equation}
\nonumber
{\bf F}_{\rm el}=-eZ{\bf E}.
\end{equation}
As for the ion drag force on the grain, ${\bf F}_{\rm dr}$, it consists of two parts arising due to the ion absorption and
ion scattering \citep{UFN,Fortov05}. For astrophysical conditions the latter provides the major contribution due to a large
value of the Coulomb logarithm, ${\mathcal L}\equiv\ln(\lambda_{\rm D}/a)\gg1$, which yields
\begin{equation}\label{ID}
{\bf F}_{\rm dr}\simeq\frac{2\sqrt{2\pi}}3m_in_iv_{T_i}{\mathcal L}\left(\frac{e^2Z}{k_{\rm B}T}\right)^2{\bf u}_i,
\end{equation}
where $n_{i}$ and ${\bf u}_i$ are the local ion density and flow velocity, respectively, and $v_{T_i}=\sqrt{k_{\rm B}T/m_i}$
is the ion thermal velocity. Furthermore, when considering a force on dust per unit volume, $\propto\int da\;(dn_d/da){\bf
F}(a)$, one should take into account that ${\bf F}_{\rm el}\propto a$ and ${\bf F}_{\rm dr}\propto a^2$ (for given ${\bf E}$
and ${\bf u}_i$). Thus, both forces are dominated by the small-size part of the MRN distribution \citep{Mathis}, i.e., the
balance per unit volume is equivalent to the balance of forces on a single grain of some effective size (near the small-size
cutoff).

The intrinsic length scale to be used further for the normalization is the ion mean free path (for a fluid description of
ions employed in this paper, it should be much smaller than the characteristic macroscopic length scale). Since the density
of neutrals is constant, we choose the ion mean free path due to collisions with neutral hydrogen,
$\ell=(\sigma_{in}n_n)^{-1}$, as the natural length scale for the normalization of coordinates and forces (as well as of the
electric field),
\begin{equation}
\nonumber
{\bf R}=\frac{\bf r}{\ell},\quad\tilde{\bf F}\equiv\frac{\ell{\bf F}}{k_{\rm B}T},\quad
\tilde{\bf E}\equiv\frac{e\ell{\bf E}}{k_{\rm B}T}.
\end{equation}
In the normalized form, the electrostatic and drag forces are
\begin{equation}
\nonumber
\tilde{\bf F}_{\rm el}=-Z\tilde{\bf E},\quad\tilde{\bf F}_{\rm dr}=ZzN{\bf U},
\end{equation}
where
\begin{equation}
\nonumber
{\bf U}\equiv\frac{{\bf u}_{i}}{\sqrt{2}v_{T_i}},\quad N\equiv\frac{n_{i}}{n_{\rm eff}},
\end{equation}
and
\begin{equation}\label{density_norm}
n_{\rm eff}=\frac{3k_{\rm B}T\sigma_{in}}{4\sqrt{\pi}{\mathcal L} e^{2}a}n_n.
\end{equation}
Inside the dust region, the balance of averaged forces per grain yields
\begin{equation}\label{balance_dust}
\tilde{\bf E}=zN{\bf U}.
\end{equation}
Thus, we derived the equilibrium equation relating the ion flow velocity and the associated electric field. The field can be
found without solving the Poisson equation, since the quasineutrality condition,
\begin{equation}\label{QN}
P=N-n,
\end{equation}
is satisfied with the very good accuracy $\sim\ell/\lambda_{\rm D}$. Here, the electron density $n_e$ is normalized by
$n_{\rm eff}$ while for the dust density $n_d$ the modified Havnes parameter $P$ is implemented:
\begin{equation}
\nonumber
n=\frac{n_{e}}{n_{\rm eff}}, \quad P\equiv\frac{Zn_{d}}{n_{\rm eff}}.
\end{equation}
Note that $P$ is different from the conventionally used Havnes parameter $P_{\rm H}=Zn_d/n_{e}$ \citep{Fortov05} -- the
latter also depends on varying plasma density and therefore is not convenient for the dimensionless analysis.

\section{Master equations for equilibrium spherical clouds: Negatively charged dust}
\label{master_eqs}

Let us derive self-consistent conditions of equilibrium for spherically-symmetric dusty clouds. The force balance for ions
should include the momentum loss due to collisions with dust. Using Eq.~(\ref{balance_dust}), we express the latter as
$-(n_{d}/n_{i})F_{\rm dr}=-(P/N)\tilde E$ (force per ion). Neglecting the ram pressure force $UdU/dR$ (since $U\ll 1$) and
taking into account the ion pressure, we get the ion force balance equation,
\begin{equation}\label{balance_ions_force}
\left(1-\frac{P}{N}\right)\tilde E-U-\frac{1}{N}\frac{dN}{dR}=0.
\end{equation}
Here $\tilde E$ is the radial component of the electric field and the term $-U$ describes the ion friction on neutrals (ion
mobility equals to unity in the employed normalization). The force balance for electrons is governed by the Boltzmann
equilibrium,
\begin{equation}\label{balance_electrons_force}
\tilde E+\frac{1}{n}\frac{dn}{dR}=0.
\end{equation}
The ion flux $NU$ is determined by the ion absorption on dust grains (found from a simple OML charging theory
\citep{Fortov05}). For a negatively charged dust we get
\begin{equation}\label{balance_ion_flux}
\frac{1}{R^{2}}\frac{d}{dR}\left(R^{2}NU\right)=-\left(1+\frac{1}{z}\right)\frac{3PN}{2{\mathcal L}}.
\end{equation}
The change of the normalized charge $z$ with the distance $R$ can be found by differentiating the OML charging equation
\citep{Fortov05},
\begin{equation}\label{charging}
\sqrt{\mu}e^{-z}n=\left[z+1-(z-1)\frac{U^2}3\right]N,
\end{equation}
where $\mu\equiv m_i/m_e~(=1840)$ is the ion-to-electron mass ratio (for a hydrogen plasma). Here we took into account that
the electron flux is thermal, while in ion flux the lowest-order (quadratic) velocity corrections are included. The latter
is necessary since all terms in $dz/dR$ are proportional to $U$.

By combining Eqs~(\ref{balance_ions_force})-(\ref{balance_ion_flux}) with the quasineutrality condition (\ref{QN}) and
differentiating Eq.~(\ref{charging}) we obtain the following set of master equations for the case of negatively charged
dust:
\begin{eqnarray}
  \frac{dN}{dR} = (zn-1)NU, \hspace{4.35cm}\label{master_ni}\\
  \frac{dn}{dR} = -znNU, \hspace{5cm}\label{master_ne}\\
  \frac{dU}{dR} = -\frac{2U}{R}-\left(1+\frac{1}{z}\right)\frac{3P}{2{\mathcal L}},\hspace{2.97cm}\label{master_ui}\\
  \frac{dz}{dR} = -\bigg\{\frac{z+1}{z+2}(zN+zn-1)\hspace{2.77cm}\nonumber\\
  \left.+\frac{z-1}{z+2}\left[\left(1+\frac{1}{z}\right)\frac{P}{{\mathcal L}}+\frac{4U}{3R}\right]\right\}U.\label{master_z}
\end{eqnarray}
The equations are written in the form which does not contain the dust size explicitly (except for the Coulomb logarithm
${\mathcal L}$, which can be considered constant for relevant astrophysical environments). Note that the ionization is
assumed to be small and therefore is neglected (the role of ionization is briefly discussed in the end of
Sec.~\ref{equilibrium_results}).

To solve the master equations it is sufficient to know the values $N_0\equiv N(0)$ and $P_0\equiv P(0)$ at the center, since
$n_0\equiv n(0)$ can be found from the quasineutrality condition. The dust charge at the center $z_0$ is found from
Eq.~(\ref{charging}),
\begin{equation}
\nonumber
\frac{e^{-z_0}}{z_0+1}\left(1-\frac{P_0}{N_0}\right)=\frac1{\sqrt{\mu}}.
\end{equation}
The flux at the center vanishes as $U(R)\to U_{0}'R$ for $R\to0$. According to Eq.~(\ref{master_ui}),
\begin{equation}
\nonumber
U_0'=-\left(1+\frac{1}{z_0}\right)\frac{P_0}{2{\mathcal L}}<0,
\end{equation}
since $P_0>0$, i.e., the plasma flux is naturally directed towards the center. Thus, the only restriction in formation of
dusty clouds is
\begin{equation}
\label{ineq}
N_0-P_0>0
\end{equation}
In what follows, we limit ourselves to the consideration of spherically-symmetric clouds.

\section{Numerical results and estimates for ISM}
\label{equilibrium_results}

\begin{figure}
\includegraphics[width=\columnwidth]{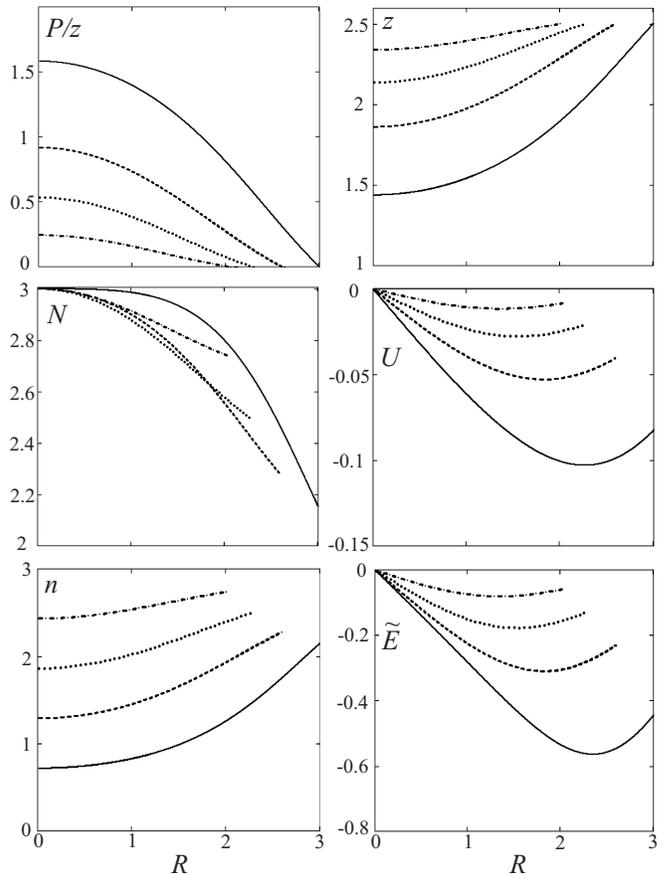}
\caption{Radial distribution of parameters in a compact cloud of negatively charged dust: The normalized
dust number density $P/z$, charge $z$, ion density $N$, drift velocity $U$,
electron density $n$, and electric field $\tilde E$ versus the distance $R$ from the center. The results are for
$N_0=3$ and four different combinations of $P_0$ and $n_0$ (related by $N_0=P_0+n_0$, represented by different lines). The
size of the cloud $R_{\rm cl}$ is determined by the condition $P(R_{\rm cl})=0$.} \label{Fig.2}
\end{figure}

\begin{figure}
\includegraphics[width=\columnwidth]{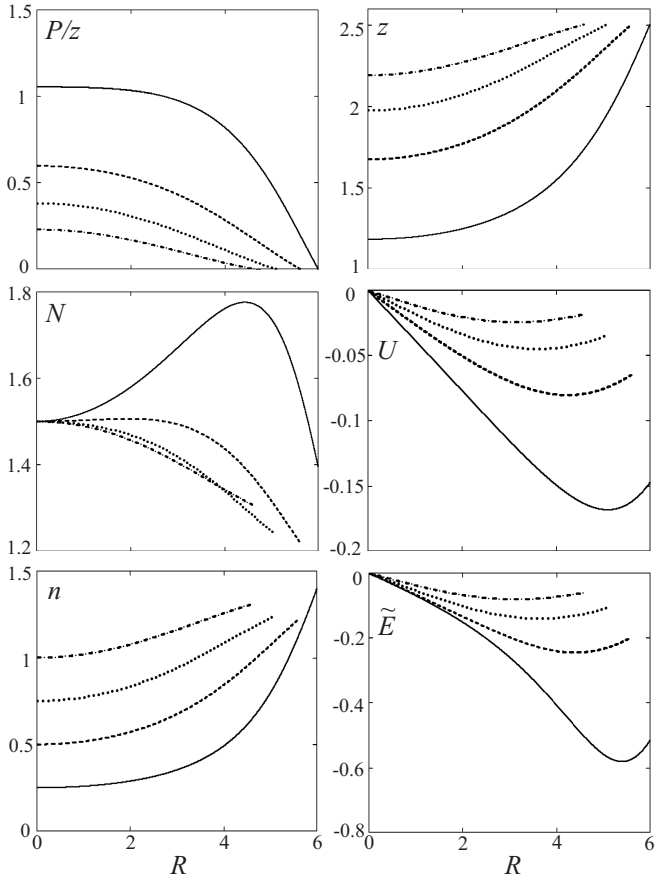}
\caption{The same as Fig.~\ref{Fig.2}, but for $N_0=1.5$. Note that the ion density profile become non-monotonous.} \label{Fig.3}
\end{figure}

We carried out calculation for various values of $N_0$ and $P_0$ limited by condition (\ref{ineq}), and did not find any
solutions other than compact dusty clouds with the size $R_{\rm cl}$ determined by $P(R_{\rm cl})=0$. A qualitative change
in spatial distribution of the cloud parameters was only found when $P_0$ approaches $N_0$. We also checked the
applicability of the hydrodynamic approach, defined by the condition that the characteristic inhomogeneity length should be
much larger than the mean free path of ions, and found that all clouds with $N_0\lesssim10$ satisfy the condition. The
calculations indicate that $R_{\rm cl}$ increases with decreasing $N_0$, implying better applicability of hydrodynamic
approach. We note that the cloud size increases approximately proportional to the Coulomb logarithm ${\mathcal L}$. All
calculations shown below were performed for ${\mathcal L}=20$.

The results are illustrated in Figs~\ref{Fig.2} and \ref{Fig.3}. One can see qualitative changes occurring in the radial
density distributions as $N_0$ decreases: While for $1.5\lesssim N_0\lesssim 10$ both dust and ion densities decrease
monotonously towards the edge (Fig.~\ref{Fig.2}), at $N_0\simeq1.5$ an ion density hump appears near the edge
(Fig.~\ref{Fig.3}). The dust number density ($P/z$) remains monotonous in the latter case, but the dust charge density ($P$)
exhibits a hump as well. The hump grows with the further decease of $N_0$, so that charges mainly concentrate at the
periphery of the cloud, resulting in the steepening of density gradients as $R\to R_{\rm cl}$.

Let us discuss implications for ISM. From Table~\ref{tab1} we see that the background ionization fraction, $n_{i,{\rm
b}}/n_n$, varies in different environments in the range from $\sim10^{-1}$ for WNM to $\sim10^{-3}$ for RN and CNM. For the
ion density inside the dust clouds we employ Eq.~(\ref{density_norm}),
\begin{equation}\label{example1}
\frac{n_i}{n_n}=\frac3{4\sqrt{\pi}{\mathcal L}}\frac{N}{z}\frac{Z\sigma_{in}}{a^2},
\end{equation}
with the momentum transfer cross section $\sigma_{in}\sim10^{-14}$~cm$^2$ (it slightly decreases with $T$ and is about the
same for collisions with atomic and molecular hydrogen; the used value corresponds to $k_{\rm B}T\sim0.1$~eV). Since the ion
density at the cloud edge is practically equal to the background density $n_{i,{\rm b}}$ (with a slight deviation is due to
finite ion velocity), from Eq.~(\ref{example1}) we conclude that Figs~\ref{Fig.2} and \ref{Fig.3} would represent typical
dusty clouds in WNM with $a\sim3\times10^{-7}$~cm, and in RN with $a\sim3\times10^{-6}$~cm. The cloud size $r_{\rm cl}\equiv
R_{\rm cl}\ell$, being of the order of a few ion-neutral mean free paths $\ell=(\sigma_{in}n_n)^{-1}$, varies in the range
between $\sim30$~AU for WNM and $\sim10^{-2}$~AU for RN.

As regards the dust density inside such clouds, it can be estimated from the relation
\begin{equation}\label{example2}
\frac{n_d}{n_n}=\frac3{4\sqrt{\pi}{\mathcal L}}\frac{P}{z}\frac{\sigma_{in}}{a^2}.
\end{equation}
Using the MRN size distribution \citep{Mathis} we obtain that the background number density of dust with size $\geq a$
follows the scaling $n_{d,{\rm b}}(a)/n_n\sim0.3A_{\rm MRN}a^{-2.5}$, where $A_{\rm MRN}\sim10^{-25}$~cm$^{2.5}$ is the
scaling constant \citep{Draine84}. For $a\sim10^{-6}$~cm, from Eq.~(\ref{example2}) we get $n_d/n_n\sim3\times10^{-4}$ at
the center, whereas the corresponding background density is $n_{d,{\rm b}}/n_n\sim3\times10^{-11}$. Hence, the dust density
inside the clouds can be increased by a factor of $\sim10^7$ with respect to the background level (note that this factor
scales as $\propto a^{0.5}$). Therefore, the total mass of the cloud $m_{\rm cl}$ is dominated by dust and can be roughly
estimated as $m_{\rm cl}\sim\frac43\pi r_{\rm cl}^3m_dn_d$, where $m_{d}=\frac43\pi a^{3}\rho_{d}$ is the mass of a dust
grain expressed via the effective grain density $\rho_{d}$. Assuming $\rho_d\sim1$~g/cm$^3$, for WNM we get $m_{\rm
cl}\sim10^{-3}$ Earth masses. The cloud mass approximately scales as $m_{\rm cl}\propto a/n_n^2$, so that the presented
estimate for WNM should be considered as the upper bound.

The optical depth of dusty clouds is expected to be determined by the Rayleigh scattering. Using the scattering cross
section \citep{LandauED} $\sigma_{\rm R}=\frac{128}{3}\pi^5a^6/\lambda^4$ (where $\lambda$ is the optical wavelength) and
employing Eq.~(\ref{example2}) we conclude that the optical depth, $\sim\sigma_{\rm R}n_dr_{\rm cl}\propto (a/\lambda)^4$,
has a very strong dependence on the dust size and is not (explicitly) related to $n_n$. For $a\sim3\times10^{-6}$~cm the
optical depth can reach values of $\sim10^{-2}$, i.e., dusty clouds are expected to be optically thin for the discussed
astrophysical environments.

To conclude this section with two notes concerning applicability of our model:

(i) The volume ionization was neglected in the above consideration. We investigated the effect of this process inside the
cloud by adding the source term $\propto n$ to the r.h.s. of Eq~(\ref{balance_ion_flux}). The analysis showed that as long
as the plasma loss on dust dominates (i.e., the source term is much smaller than the absorption term) the overall
distributions as well as the size of the cloud remain practically unchanged.

(ii) For the applicability of fluid approach, the mean free path of ions (which is determined by collisions with neutrals
and dust) must be much smaller than the characteristic length of density inhomogeneity. This requirement results in the
condition,
\begin{equation}\label{fluid_condition}
\frac{d\ln N}{dR}\ll 1+\frac{zP}{\sqrt{2}},
\end{equation}
which is always satisfied in the central region, but can be violated near the cloud edge. For the examples shown in
Figs~\ref{Fig.2} and \ref{Fig.3}, the r.h.s. of Eq.~(\ref{fluid_condition}) at $R=R_{\rm cl}$ is 5--10 times larger than the
l.h.s. and, hence, the use of fluid approach is well justified.

\section{Stability of clouds: Spherical perturbations}
\label{stability}

As we pointed out in Sec.~\ref{equilibrium_results}, the fluid approach is well applicable for calculating the clouds
illustrated in Figs~\ref{Fig.2} and \ref{Fig.3}. Now, the principal question is under which conditions the clouds are
stable, in particular -- whether the clouds with non-monotonous charge density distributions shown in Fig.~\ref{Fig.3} can
be unstable.

To address this question, let us consider eigenmodes of the equilibrium clouds. We shall assume small (linear) deviations
from steady state, and restrict ourselves to the analysis of spherical perturbations (which are presumably the most
``dangerous'' for the stability). In the dimensional form, we consider perturbations of the electron and ion densities,
$\delta n_{e,i}(r)$, ion velocity $\delta u_i(r)$, dust charge $\delta Z(r)$, and dust velocity $u_{d}(r)$ (we do not write
$\delta$ in front of $u_d$ since it is zero in equilibrium), all proportional to $e^{-i\omega t}$.

The new variable $u_{d}$ as well as the frequency $\omega$ and mass of a dust grain $m_d$ are normalized in such a way that
both the dust inertia force, $-i\omega m_du_d$, and the friction force due to dust interaction with the ambient gas
\citep{Fortov05,Purcell}, $\frac{8\sqrt{2\pi}}{3}a^2m_nn_nv_{T_n}u_d$, are written in the simplest dimensionless form. This
yields
\begin{equation}
\nonumber
V\equiv\frac{u_{d}}{u_{\rm eff}},\quad \Omega\equiv\frac{\ell\omega}{u_{\rm eff}},\quad
M\equiv \frac{e^{2}}{ak_{\rm B}T}\frac{m_{d}u_{\rm eff}^{2}}{k_{\rm B}T},
\end{equation}
where
\begin{equation}
\nonumber
u_{\rm eff}=\frac{3k_{\rm B}T\sigma_{in}}{8\sqrt{2\pi}e^{2}a}v_{T_n}.
\end{equation}
Here $v_{T_n}=\sqrt{k_{\rm B}T/m_n}~(\simeq v_{T_i})$ is the thermal velocity of neutrals.

The analysis of spherically-symmetric perturbations is relatively straightforward. In the steady state, the balance of drag
and electric field forces has the form of Eq.~(\ref{balance_dust}). By taking into account the friction and inertial forces
in the momentum equation for dust, we obtain the field perturbation,
\begin{equation}
\nonumber
\delta \tilde E=\delta(zNU)-\frac{1}{z}(1-iM\Omega)V,
\end{equation}
whereas the perturbation of the dust charge is found from charging equation (\ref{charging}),
\begin{equation}
\nonumber
\delta z=-\frac{z+1}{z+2}\left(\frac{\delta N}{N}-\frac{\delta n}{n}\right)+\frac23\left(\frac{z-1}{z+2}\right)U\delta U.
\end{equation}
These relations are used in equations for perturbations of the ion and electron densities derived from
Eqs~(\ref{balance_ions_force}) and (\ref{balance_electrons_force}),
\begin{eqnarray}
  \frac{d\delta N}{dR} &=& \delta(\tilde En-NU),\label{perturb_ni} \\
  \frac{d\delta n}{dR} &=& -\delta(\tilde En).\label{perturb_ne}
\end{eqnarray}
The equation for the ion velocity perturbation follows from Eq.~(\ref{master_ui}),
\begin{equation}\label{perturb_ui}
\frac{d\delta U}{dR}=-\frac2{R}\delta U-\frac3{2{\mathcal L}}\left(1+\frac1{z}\right)\delta P
+\frac{3P}{2{\mathcal L} z^2}\delta z,
\end{equation}
while for the dust velocity perturbation we use the continuity equation,
\begin{eqnarray}\label{perturb_ud}
\frac{dV}{dR}=i\Omega\left(\frac{\delta P}{P+\delta P}-\frac{\delta z}{z}\right)\hspace{2.7cm}\nonumber\\
-\left(\frac{2}{R} +\frac{dP/dR}{P+\delta P}-\frac{dz/dR}{z}\right)V,
\end{eqnarray}
where $P$ is the equilibrium value of the Havnes parameter and $\delta P$ is its perturbation,
\begin{equation}
\nonumber
\delta P=\delta N -\delta n.
\end{equation}

Generally, Eq.~(\ref{perturb_ud}) requires special consideration near the cloud edge, since it becomes essentially nonlinear
when $R\to R_{\rm cl}$. On the other hand, Figs~\ref{Fig.2} and \ref{Fig.3} show that $P(R)$ is a rather steep function near
the edge, so that the nonlinearity is only important in a close proximity of $R_{\rm cl}$. Thus, to a first approximation
one can neglect $\delta P$ in the denominator of Eq.~(\ref{perturb_ud}) for the analysis of eigenmodes, and then
Eqs~(\ref{perturb_ni})-(\ref{perturb_ud}) can be presented in the following matrix form:
\begin{equation}\label{perturbed_eq}
\frac{d{\bf X}}{d{\bf R}}=\textsf{A}\cdot{\bf X},
\end{equation}
where ${\bf X}=(\delta N,\delta n,\delta U,V)^{\rm T}$ is the perturbation vector and the elements of matrix $\textsf{A}$
are given in Appendix~\ref{app2}.

Equation~(\ref{perturbed_eq}) requires four boundary conditions. The first two obvious conditions are $\delta U=0$ and $V=0$
at the center. By substituting $\delta U(R)\to \delta U_{0}'R$ in Eq.~(\ref{perturb_ui}) and $V(R)\to V_0'R$ in
Eq.~(\ref{perturb_ud}) for $R\to 0$, we find
\begin{eqnarray*}
  \delta U_{0}' = -\frac1{2{\mathcal L}}\bigg[\left(1+\frac{1}{z_0}\right)\delta P_0\hspace{2.7cm}\nonumber\\
  +\frac{z_0+1}{z_0^2(z_0+2)}P_0\left(\frac{\delta N_0}{N_0}-\frac{\delta n_0}{n_0}\right)\bigg],\\
  V_0' = \frac{i\Omega}{3}\left[\frac{\delta P_0}{P_0}+\frac{z_0+1}{z_0(z_0+2)}\left(\frac{\delta N_0}{N_0}-
  \frac{\delta n_0}{n_0}\right)\right],
\end{eqnarray*}
(where $\delta P_0=\delta N_0 -\delta n_0$). The other two conditions follow from the assumptions that (i) the number of
grains is conserved in perturbations,
\begin{equation}
\int_{0}^{R_{\rm cl}}dR\;R^{2}\left(\frac{\delta P}{z}-\frac{P\delta z}{z^{2}}\right)=0,
\end{equation}
and (ii) the external plasma flux at the cloud edge is not affected by perturbations,
\begin{equation}
(U\delta N+N\delta U)|_{R_{\rm cl}}=0.
\end{equation}

\subsection{Examples of eigenmodes}

The numerical solution of Eq.~(\ref{perturbed_eq}) was carried out by separating the real and imaginary parts and
representing the eigenmodes in the form $\Omega =\Omega_{\rm re} +i\Omega_{\rm im}$. The most important conclusion of our
analysis is that the eigenmodes remain stable for all combinations of parameters studied in Sec.~\ref{equilibrium_results}.
For a given pair of $P_0$ and $N_0$ we get a finite number of eigenmodes with increasing $\Omega_{\rm re}$ and practically
constant $\Omega_{\rm im}$ (the latter is primarily determined by the gas friction). The modes also depend on the
dimensionless mass of a dust grain, which can be presented in the following form:
\begin{equation}
\nonumber
M=\frac{9}{128\pi}\frac{Zm_d}{zm_n}\left(\frac{\sigma_{in}}{a^2}\right)^2.
\end{equation}
Since $Z$ is roughly proportional to the size $a$ and $m_d\propto a^3$, the dimensionless mass is practically independent of
$a$ and varies in the range $M=100-1000$ for different ISM.

Let us illustrate the eigenmodes for a spherical cloud corresponding to Fig.~\ref{Fig.2}. By setting the dust density
$P_0=2.28$ (solid line) and mass $M=100$, we obtain a discrete series with slowly increasing $\Omega_{\rm re}$ (varying in
the range between 0.107 and 0.633 for the first six modes) and constant $\Omega_{\rm im}=-0.005$.
This represents weakly damped dust-acoustic wave modes sustained in a cloud with inhomogeneous distribution of dust and
ions. One can roughly estimate the frequency in dimensional units as $\omega_{\rm re}\sim 2\pi C_{\rm DA}/r_{\rm cl}$, where
$C_{\rm DA}=Z\sqrt{(n_d/n_i)k_{\rm B}T/m_{d}}$ is the dust-acoustic speed \citep{Fortov05}. In the dimensionless units this
yields $\Omega_{\rm re}\sim(2\pi/R_{\rm cl})\sqrt{zP/MN}$, which is indeed close to the lowest value of $\Omega_{\rm re}$
from the calculated series.

\section{Dusty clouds formed by positively charged grains}
\label{positive}

Generally, dust charges in a cosmic environment are determined by a balance of plasma currents absorbed on grains (resulting
in negative charges) and the photoemission currents induced by cosmic UV radiation (tending to charge grains positively)
\citep{Draine09,Draine11}. The analysis of dusty clouds becomes rather complicated in this case. Instead of solving the
problem self-consistently, here we study the regime when the UV radiation dominates and dust acquires large {\it positive}
charges $eZ$. Such situation occurs when the electron photoemission current from the grain exceeds the current of
surrounding electrons on the grain \citep{Khrapak99}, i.e., when $\gamma J_{\rm UV}\gg n_e\sqrt{k_{\rm B}T/m_e}$, where
$J_{\rm UV}$ is the UV photon flux and $\gamma$ is the yield of photoelectrons (for simplicity, we assume the photoelectron
temperature to be equal to the plasma temperature $T$). In this regime the normalized grain charge $z$, being written in the
form of Eq.~(\ref{norm_charge}), has a logarithmic dependence on $\gamma J_{\rm UV}$ and therefore depends on external
conditions (in contrast to the case of negatively charged grains considered above).

\begin{figure}
\includegraphics[width=\columnwidth]{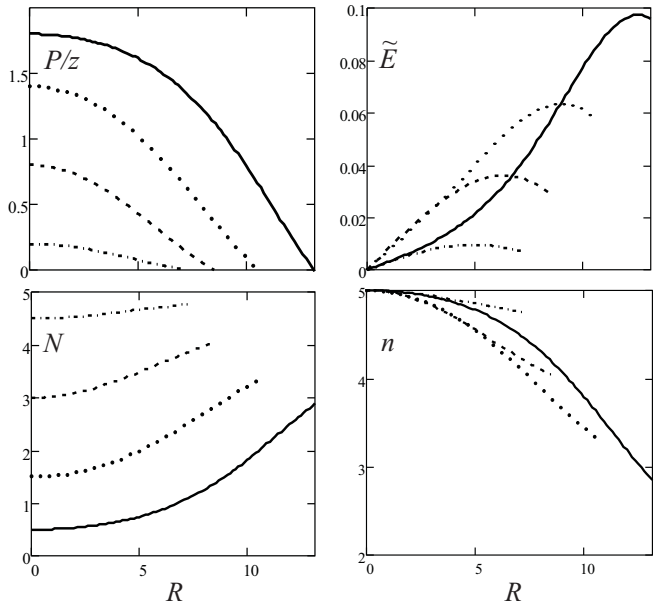}
\caption{Radial distribution of parameters in a compact cloud of positively charged dust: The normalized
dust number density $P/z$, electric field $\tilde E$, ion density $N$, and electron density $n$ versus the distance $R$ from
the center. The results are for $n_0=5$ and four different combinations of $P_0$ and $N_0$ (related by $n_0=P_0+N_0$,
represented by different lines). The normalized charge $z=2.5$ is fixed. The size of the cloud $R_{\rm cl}$ is determined
by the condition $P(R_{\rm cl})=0$.} \label{Fig.4}
\end{figure}

In order to analyze master equations for equilibrium clouds formed by positively charged dust, let us for simplicity fix the
value of the normalized charge $z~(=2.5)$. Then we can replace $z\to-z$ in the above equations and expect that the electric
field reverts the sign as well, $E\to-E$. In this case Eq.~(\ref{balance_dust}), which determines the force balance on dust,
is still valid. Furthermore, we replace $P\to-P$, so that the quasineutrality condition, Eq.~(\ref{QN}), is modified to
\begin{equation}\label{QN+}
P=n-N.
\end{equation}
Finally, using the OML ion current on positively charged grains \citep{Fortov05}, from
Eqs~(\ref{master_ni})-(\ref{master_ui}) we get the following set of master equations:
\begin{eqnarray}
  \frac{dN}{dR} &=& -(zn+1)NU, \hspace{4.35cm}\label{master_ni+}\\
  \frac{dn}{dR} &=& znNU, \hspace{5cm}\label{master_ne+}\\
  \frac{dU}{dR} &=& -\frac{2U}{R}-\frac{}{}\frac{3P e^{-z}}{2{\mathcal L}z}.\hspace{2.97cm}\label{master_ui+}
\end{eqnarray}
The numerical solution of the equations, obtained for the parameters similar to those in Fig.~\ref{Fig.2}, is presented in
Fig.~\ref{Fig.4}. One can see that clouds of positively charged dust become several times larger, while the dust density
profile keeps the same form. The electric field reverses the sign (and also becomes weaker), which ensures the force balance
in Eq.~(\ref{balance_dust}) and modifies distributions of the ion and electron densities, in such a way that the form of
$n(R)$ resembles that of $N(R)$ in Fig.~\ref{Fig.2} and vise versa. The ion velocity profile (not shown) remains similar to
that in Fig.~\ref{Fig.2}.

Note that in order to get a universal dependence on $z$, one can simply re-scale variables in
Eqs~(\ref{QN+})-(\ref{master_ui+}). For instance, the solution for arbitrary $z$ is obtained from that for $z=1$ by
employing the re-scaling $z^{-1}\{n,N,P\}\to\{n,N,P\}$, $\alpha^{-1} U\to U$, and $\alpha R\to R$, where
$\alpha=ze^{\frac12(z-1)}$.

Thus, we conclude that positively charged dust can also form compact clouds. This suggests that the mechanism of
self-confinement associated with plasma absorption on grains can indeed be generic and independent on the dominating
charging process.

\section{Compact dusty clouds and the small-scale structure of the diffuse ISM}
\label{diffuse}

So far, in our consideration we neglected feedback to the surrounding medium. In particular, we assumed that properties of
neutral gas remain unaffected in the presence of dusty clouds. However, the dust density inside the clouds is increased by
many orders of magnitude. Therefore, the clouds may in fact play a role of ``condensation seeds'' for the gas: Concentrated
dusty cores may effectively cool down the surrounding gas due to efficient gray-body radiation and, hence, cause the gas
density to increase due to compression by the hotter surrounding. Thus, one might expect a formation of compact gaseous
``nuggets'' whose equilibrium is determined by the balance of the total dust radiation and the diffusive heat flux from
outside.

To demonstrate the feasibility of this process, let us first evaluate how effective the thermal coupling between dusty
clouds and the surrounding gas would be. One can expect that the gas temperature inside the cloud is close to the dust
temperature provided the mean free path of atoms/molecules due to collisions with dust, $\ell_{nd}=(\pi a^2 n_d)^{-1}$, does
not exceed the cloud size $r_{\rm cl}$. As we see from Figs~\ref{Fig.2}-\ref{Fig.4}, the latter varies between a few and a
few dozens of the ion mean free path $\ell=(\sigma_{in}n_n)^{-1}$. Taking this into account and also using
Eq.~(\ref{example2}) we obtain that the ratio $r_{\rm cl}/\ell_{nd}$ is practically a universal constant,
\begin{equation}
\nonumber
\frac{r_{\rm cl}}{\ell_{nd}}=\frac{3\sqrt{\pi}}{4{\mathcal L}}\frac{P}{z}R_{\rm cl},
\end{equation}
which is of the order of unity for any astrophysical environment. Thus, the gas temperature inside dusty clouds should
always be close to the dust temperature.

Next, we estimate the efficiency of the radiative dust cooling. The cooling time $\tau_{\rm rad}$ for an individual particle
can be roughly obtained from the following balance: $4\pi a^2\sigma_{\rm SB}T_d^4\sim \frac43\pi a^3\rho_d c_d k_{\rm
B}T_d/\tau_{\rm rad}$, where $\sigma_{\rm SB}$ is the Sefan-Boltzmann constant, $c_d$ is the specific heat of a grain
material (whose effective mass density is $\rho_d\sim1$~g/cm$^3$), and $T_d$ is the (effective) dust temperature. Assuming
$T_d\sim 30$~K and $a\sim10^{-6}$~cm we get $\tau_{\rm rad}\sim10$~s. This timescale should be compared with the
characteristic heat diffusion time $\tau_{\rm diff}\sim r_{\rm cl}^2/(v_{T_n}\ell_{nn})$, where $\ell_{nn}=
(\sigma_{nn}n_n)^{-1}$ is the neutral mean free path. Taking into account that for hydrogen $\sigma_{nn}\sim0.1\sigma_{in}$,
we get $\tau_{\rm diff}\sim r_{\rm cl}/v_{T_n}\sim 1$~year for $r_{\rm cl}\sim1$~AU. We conclude that the radiation is a
very efficient cooling mechanism and, hence, $T_d$ can be much smaller than the ambient temperature $T$.

Finally, we obtain the temperature profile of the gas around a dusty cloud, $T_n(r)$, by employing the conservation of the
total diffusive heat flux, $r^2n_nv_{T_n}dT_n/dr=$~const. By solving this with the boundary conditions $T_n(r_{\rm cl})=T_d$
and $T_n(\infty)=T$ and under the assumption of a constant pressure, $n_n(r)T_n(r)=$~const, we get that the size of the
``cold atmospheres'' around dusty clouds is always about a few $r_{\rm cl}$. For example, the size of gaseous nuggets
created by dusty clouds in WNM can reach a few hundreds of AU, while the gas density inside such objects can be increased by
a factor of $\sim T/T_d\sim100$ (for $T_d\sim30$~K).

\section{Discussion and conclusions}
\label{conclusions}

We showed that dust grains in various ISM can self-organize themselves into equilibrium compact spherical clouds. The
formation of such clouds is caused by the ion ``shadowing'' forces \citep{MT00,BT01,TW03}. This triggers the ``shadowing''
instability which resembles the Jenans instability and originates from the generic mechanism of plasma absorption on dust
grains. The confinement stabilizing equilibrium clouds is generated by the inward plasma flux which, in turn, is created due
to the plasma absorption on dust. Let us summarize the major characteristics of compact dusty clouds:

(i) After proper normalization, the cloud is characterized by a combination of two global parameters -- the normalized ion
and dust densities at the center. The former is determined by the background plasma density in a given astrophysical
environment, the latter is a free parameter whose variation is only limited by condition (\ref{ineq}).

(ii) For typical astrophysical conditions, the size of compact clouds is below $\sim30$~AU. The dust density inside the
clouds can exceed the background density by about 7 orders of magnitude, and their total mass can be as large as
$\sim10^{-3}$ Earth masses. At the same time, the optical depth of the clouds (determined by the Rayleigh scattering)
remains much smaller than unity.

There have been a few reports on the existence of small-scale objects observed in different astrophysical environments,
whose origins remain unclear. In particular, ultra-high resolution observation of interstellar absorption
\citep{BK05,Lauroesch,S13,C13} have provided convincing evidence that the diffuse ISM is structured on scales below 1 pc
down to dozens of AU, indicating the presence of tiny, distinct, dense and cold HI cloudlets, the origin of which is not
understood up to now. As sketched in Sec.~\ref{diffuse}, compact dusty clouds might trigger formation of such cloudlets.
However, the details of their growth in the turbulent diffuse ISM are certainly more complex than outlined by the simple
hydrostatic model of Sec.~\ref{diffuse}. We plan to investigate this interesting process in detail in a subsequent paper.
Also, the motion of compact clouds near the Galactic center has been recently detected \citep{Gillessen}. Some properties of
these clouds (e.g., small optical depth, size $\sim100$~AU, and the mass smaller the Earth mass) are similar to those
predicted by our model. The UV radiation near the Galactic center is very strong, so that dust is positively charged and
characteristics of the resulting clouds should be similar to those presented in Sec.~\ref{positive}. On the other hand, the
tidal forces exerted on clouds in this region should be so strong (see Appendix~\ref{app1}) that their equilibrium is highly
questionable and requires further analysis.

In conclusion, we would like to note that dust in ISM is extremely polydisperse and, hence, different clouds are expected to
contain grains of different sizes. Understanding the size distribution of dusty clouds and their morphology would be a very
important problem to address. Another very interesting topic is to investigate coupling of dusty clouds (and the surrounding
atmosphere) to the dynamics of the ambient ISM. Since the characteristic size of dusty clouds is expected to be much smaller
that the cutoff scale of the MHD turbulence, one can use results of this paper as the input for such analysis.


\begin{acknowledgements}
This work, performed in the Max Planck Institute for Extraterrestrial Physics (MPE) and General Physics Institute RAS, is
supported by the Max Planck Society. One of the authors (V.N.T.) thanks for the hospitality while staying at MPE.
\end{acknowledgements}

\appendix

\section{A. Condition to neglect tidal forces}
\label{app1}

The formation of dusty clouds in a close proximity to stars or black holes requires more careful analysis due to the
presence of gravitational forces neglected in this paper. Below we estimate the condition when the resulting tidal force is
small in comparison with the ion drag force [Eq.~(\ref{ID})].

For a cloud of the radius $r_{\rm cl}$, comprised of dust grains of size $a$ and located at the distance $d_{\rm star}$ from
a star/black hole of the mass $m_{\rm star}$, the tidal force is negligible when
\begin{eqnarray*}
\left(\frac{r_{\rm cl}}{\rm 1~AU}\right)\lesssim {\mathcal L}\left(\frac{m_{\odot}}{m_{\rm star}}\right)
\left(\frac{d_{\rm star}}{\rm 100~AU}\right)^3\left(\frac{1~\mu{\rm m}}{a}\right)\left(\frac{1~{\rm g}/{\rm cm}^{3}}{\rho_{d}}\right)
\left(\frac{n_{i}}{10~{\rm cm}^{-3}}\right)\left(\frac{k_{\rm B}T}{1~{\rm eV}}\right).
\end{eqnarray*}
For instance, assuming $m_{\rm star}=3\times10^6~m_{\odot}$ (corresponds to a massive black hole of our Galaxy), $r_{\rm
cl}=3$~AU, $a=10^{-6}$~cm, $\rho_d=1$~g/cm$^3$, $n_i=0.3$~cm$^{-3}$, and $k_{\rm B}T=0.1$~eV, we conclude that the gravity
can be neglected when $d_{\rm star}\gtrsim10^4$~AU.

\section{B. Elements of the dynamical matrix}
\label{app2}

The dynamical matrix $\textsf{A}$ in Eq.~(\ref{perturbed_eq}) has the following elements:
\begin{eqnarray*}
A_{\delta N,\delta N}=\left(\frac{z^{2}+z-1}{z+2}n-1\right)U, \quad
A_{\delta N,\delta n}=\frac{z^2+3z+1}{z+2}NU,\hspace{6cm}\\
A_{\delta N,\delta U}=(zn-1)N, \quad
A_{\delta N,V}=-\frac1{z}(1-iM\Omega)N,\\
A_{\delta n,\delta N}=-\frac{z^{2}+z-1}{z+2}nU, \quad
A_{\delta n,\delta n}=-\frac{z^2+3z+1}{z+2}NU,\hspace{6.8cm}\\
A_{\delta n,\delta U}=-znN, \quad
A_{\delta n,V}=\frac1{z}(1-iM\Omega)n,\\
A_{\delta U,\delta N}=-\frac3{2{\mathcal L}}\left(1+\frac{1}{z}\right)\left[1+\frac1{z(z+2)}\frac{P}{N}\right], \quad
A_{\delta U,\delta n}=\frac3{2{\mathcal L}}\left(1+\frac{1}{z}\right)\left[1+\frac1{z(z+2)}\frac{P}{n}\right],\hspace{1.95cm}\\
A_{\delta U,\delta U}= -\frac{2}{R}+\frac{z-1}{z^{2}(z+2)}\frac{PU}{{\mathcal L}}, \quad
A_{\delta U,V}=0,\\
A_{V,\delta N}= i\Omega\left[\frac{1}{P}+\frac{z+1}{z(z+2)}\frac1{N}\right], \quad
A_{V,\delta n}=-i\Omega\left[\frac{1}{P}+\frac{z+1}{z(z+2)}\frac1{n}\right],\hspace{5.1cm}\\
A_{V,\delta U}=-\frac{2i\Omega}3\frac{z-1}{z(z+1)}U, \quad
A_{V,V}=-\frac{2}{R}-\frac{1}{P}\frac{dP}{dR}+\frac{1}{z}\frac{dz}{dR}.
\end{eqnarray*}


\end{document}